\documentclass[preprint,showpacs,preprintnumbers,amsmath,amssymb,floatfix]{revtex4}
\usepackage{graphicx}
\usepackage{dcolumn}
\usepackage{bm}
\usepackage{longtable}
\begin{document}

\title{ Role of semi core levels in determining the band-gaps of semiconductors : First-principles calculations with model Hamiltonians}
\author{Roby Cherian$^{1}$, Priya Mahadevan$^{1}$ and Clas Persson$^{2}$}
\affiliation{$^1$ S.N. Bose National Centre for Basic Sciences, JD-Block, Sector III,Salt Lake, Kolkata-700098, India \\
$^2$ Department of Materials Science and Engineering, Royal Institute 
of Technology, SE-10044, Stockholm, Sweden \\}
\date{\today}

\begin{abstract}
First principle calculations based on LDA/GGA approximation for the exchange
functional underestimate the position of the semi core 3$d$ levels in Ga$X$
($X$=N, P and As) semiconductors. A self-interaction correction scheme within the LDA+U/GGA+U 
approximation is found to be sufficient to correct this discrepancy. 
A consequence of this
correction is that the bandgap ($E_g$) of the semiconductors also improves. 
The belief has been that the bandgap correction comes from
modified semi core-valence interaction.
We examine this often used approximation
in great detail and find that although bandgap changes as large as 0.63 eV
for GaAs, 0.42 eV for GaP and 0.46 eV for GaN are obtained within this 
approach for $U$= 20 eV on the Ga $d$ states, only 0.1 eV, 0.1 eV and 0.15 eV for GaAs, GaP and GaN arise from
semi core-valence interaction.  
As $U$ is increased, the bandgap keeps improving. We trace this effect primarily to 
the interaction of the Ga 4$d$ states in the conduction band with the anion $p$ states.

\end{abstract}

\maketitle

{\bf I. Introduction}

Enormous progress has been made over the years in the development of
realistic theories of materials starting from a first-principles approach \cite{real-th}.
A popular method in this direction is the Kohn-Sham formalism of density 
functional theory. The Hohenberg-Kohn theorem states \cite{kohn} that the ground state
energy can be written as an exact functional of the density. Unfortunately,
the exact form of the functional is not known and approximations such as
the local density approximation (LDA) \cite{lda,self-inter} 
as well as the generalised gradient 
approximation (GGA) \cite{ggapw91,ggapbe} have been used quite successfully in describing the ground 
state properties of a wide variety of systems. There are however limitations 
of this method when one attempts to calculate the excited state properties
such as the bandgap. The Kohn-Sham band-gaps of semiconductors are smaller 
than the experimental values. Even if one ignores the fact that the 
Hohenberg-Kohn theorem is a theory of the ground state, there are 
limitations coming from the form of the exchange correlation functional. The Kohn-Sham (KS)
band-gap E$_g^{KS}$ is related to the fundamental band-gap of the solid E$_g$
by the relation $E_g=E_g^{KS}+\Delta_{xc}+\Delta_{U}$ where $\Delta_{xc}$ is the 
discontinuity in the exchange correlation functional \cite{exch_corr}
when an electron is added to the system. 
and $\Delta_{U}$ has the corrections due to correlation effects.
This discontinuity is lacking in the local density
approximation resulting in the equality of the Kohn-Sham and the 
fundamental bandgap. The second, separate issue is the effect of the 
approximation of the exchange functional which arises from a spurious
self-interaction error \cite{self-inter}. In addition electron 
correlation effects are underestimated in density functional theory which leads
to the incorrect placement of the levels. For a more detailed discussion
of the pros and cons of density functional theory see Ref~\cite{rmartin}.

For $s-p$ bonded semiconductors, the GW approximation \cite{gw} 
has been enormously
successful in improving the calculated bandgap. In this method, a 
partially or fully self-consistent solution of the Dyson equation is used 
to determine the self-energy which provides a correction to the KS-LDA
bandgap. Another contributory factor to the gap error is the cation
$d$ position in $s-p$ semiconductors \cite{semicore}, especially when these levels are
relatively shallow. The $d$ levels are not deep enough to be 
chemically inert and so they interact with the valence band states
leading to a reduction of the bandgap. A cheap and rather inexpensive
method that has been used in the literature \cite{gapcorr,chris} for correcting for the bandgap
has been the LDA+U/GGA+U method \cite{ldapu1,ldapu2}. Within this scheme a potential is introduced
on the valence / semi core states which is dependent on the atom and angular 
momentum projected density operator. This pushes the semi core /valence states
to deeper energies and therefore modifies the valence band maximum (VBM)
position as a result of the modified semi core-valence / valence interaction.

In this work we have performed a detailed analysis of the GGA+U approach for 
correcting the bandgap of semiconductors in which a $U$ is 
applied to the cation $d$ states. We also studied the 
changes in the band structure and bandgap as a function of $U$.
Aligning all the states with respect to the anion 1$s$ level for 
that $U$ value, we find 
that at $\Gamma$ point, apart from movement of the Ga 3$d$ states, 
the changes are primarily at the valence band 
maximum. All other states largely remain unshifted. As the semi core 
states are pushed to deeper energies, there would be
a corresponding reduction in the coupling between the semi core and 
valence states. Hence for largish values of the $U$, the VBM should stop
moving. This however does not happen. We examined if this was a consequence 
of the fact that the potential is angular momentum dependent, but does not distinguish
between the 'principal quantum number' of the state involved. Invoking a
description of the eigenfunctions in terms of an atomic description, for GaAs,
the potential applied on the Ga $d$ states does not distinguish between 
Ga 3$d$ and Ga 4$d$ states. 
There is a small admixture of the unoccupied Ga 4$d$ states in the 
ground state charge density. The applied potential which depends on the 
Ga $d$ density for a given value of $U$, pulls the Ga 4$d$ states 
to lower energies where they interact more strongly with the As $p$ states, 
therefore increasing the gap. This effect can be controlled by varying the 
muffin tin radius (RMT) of the sphere about Ga in which the potential is 
applied. A decreased muffin tin radius would have a reduced Ga 4$d$ component 
and therefore modifies the potential.
{\it However even after reducing the Ga 4$d$ component}, 
the gap keeps increasing although the 
changes are small. These changes we speculate could arise from modified screening 
of the $p$  states as the $d$ states are moved deeper in energy or other 
modifications in the electronic structure arising from  charge reorganization 
within the Ga states. For the first time we clarify how the bandgap is modified 
by a $U$ on the semi core $d$ states. We also provide the band offsets between 
the $U$=0 and finite $U$ results for the GaX. These  show that the self-interaction error 
due to the incorrect placement of the Ga $d$  states changes both the valence 
band offset as well as the conduction band offset. Similar conclusions were 
found earlier by Janotti et al. \cite {chris}.

{\bf II. Methodology}

We have carried out ab-initio calculations within the generalised 
gradient approximation of density functional theory. A full potential 
linearized augmented plane wave implementation in the WIEN2K code \cite{wien2k}
was used by us in our calculations. We considered the systems GaN, GaP and GaAs
and evaluated the electronic structure at the experimental lattice \cite{lattc}
constants of 4.52, 5.45 and 5.65 $\AA$ respectively. Electronic structure 
calculations were performed for different muffin tin radii and the reasons for this 
will be discussed in the text. Unless otherwise stated, the results disscussed 
will be for RMT= 2.3, 2.2 and 1.8 a.u for Ga in GaAs, GaP and GaN respectively.
The radii for As, P and N were kept fixed at 2.3, 2.0 and 1.8 a.u. 
In the calculations $G_{MAX}$ = 12 and RK= 7. 
The number of k-points used was 64 (4x4x4) for the self-consistency, while 
a gamma-centered grid of 8x8x8 was considered in the evaluation of the Density of states. 
The tetrahedron method of integration was used for the DOS calculation.  
The GGA PW91 approximation \cite{ggapw91} to the exchange functional was used. 
An additional potential within the GGA+U \cite{ldapu2} 
formalism was introduced on the  Ga $d$ states, and the changes in the electronic 
structure was calculated as a  function of $U$.
Further analysis has been carried out within a tight-binding model (discussed
later) found to be adequate to give a good description of the electronic 
structure. This enables us to estimate bandgap shifts coming from various 
interactions. In addition plane wave pseudopotential calculations \cite{vasp}, 
have been carried out using PAW potentials \cite{paw}. A gamma centered grid 
of 8x8x8 k-points and GGA-PW91 approximation for the exchange have been used.
Superlattices of the form $(GaX)_{5}/(Ga(U)X)_{5}$ were constructed to 
determine the valence and conduction band offsets introduced by $U$, measured 
with respect to the $U$=0 result. The electronic structure of the superlattices 
was carried out using a plane wave pseudopotential implementation within 
VASP \cite{vasp}. A k-points grid of 6x6x1 was used. The positions of the 
atoms making up the superlattice were fully optimized.

{\bf III. Results and discussion}

In Fig. 1 we have plotted the $s$, $p$, $d$ projected 
density of states for $U$= 0, 5 and 10 eV
for GaAs. As expected the Ga 3$d$ states are pushed deeper into 
the valence band as $U$ is varied. The gross features of the valence band
however remain unchanged. Apart from the bandgap underestimation in these
calculations that we discussed earlier, another source of error in these
calculations is the underestimation in the position of the
semi core levels. This is a consequence of the overbinding of states that
one finds in LDA-type calculations \cite{semicore}. 
Experimentally \cite{expt}, these levels
are found at -17.0, -18.8 and -18.9 eV inside the valence band 
for GaN, GaP and GaAs.

There are however some striking effects on the bandgap as $U$ is varied.
Considering the example of GaAs, one examines the variation in the 
bandgap as the position of the semi core states are varied (Fig. 2).
The bandgap increases from 0.44 eV to 0.71 eV as $U$ is increased from 
0 to 9 eV. The first question is why does
the bandgap change as the Ga 3$d$ states are pushed deeper into the 
valence band. This is illustrated schematically in Fig. 3 for the 
symmetries involved at the $\Gamma$ point. The energy splitting of the
Ga 3$d$ states in the point ion limit is shown in the left panel. One has
$e$ below $t_2$. The $t_2$ states on the As interact with the $t_2$ 
states of Ga 3$d$ and hence form bonding and antibonding states. The 
VBM position in the absence of Ga 3$d$-As $p$ interaction would have 
been where the unperturbed As $p$ derived $t_2$ states are. The 
interaction leads to a reduction in the bandgap of the semiconductor. 
Now as the Ga 3$d$ states are pushed deeper into the valence band, one
expects the VBM to stop moving at some point where the interaction 
strength becomes negligibly small.

In order to understand this further we have considered a minimal tight-binding
model with $s$,$p$ and $d$ states in the basis for both Ga and As, P. 
In the case of GaN only $s$,$p$ states were included on N. Nearest 
neighbor interactions between the anion and cation as well as next neighbor
interactions between anion pairs and cation pairs have been considered.
The comparisons of the band dispersions obtained for GaAs, GaP and GaN from
VASP and from TB fitting for the various $U$ values are given in Fig.4. The
parameters entering the TB Hamiltonian (Table I) have been
extracted by fitting the first principle results by a least square error
minimization process. The parameters are similar to values present
in the literature \cite{ddtb}. We use the values of the interactions strengths
estimated between the semi-core and valence states to estimate the renormalization
of the bandgap due to these interactions. This is found to be just 0.1 eV for GaAs
and GaP and 0.15 eV for GaN.

Now as the Ga 3$d$ states are pushed deeper into the valence band, one
expects the VBM to stop moving at some point where the interaction
strength becomes negligibly small.
We have moved the Ga 3$d$ states to absurdly deep energies using say 
a $U$ of 20~eV where we expect the interaction to go to zero. The VBM 
however does not stop moving. 
A possible reason for this is the following and is shown schematically in Fig. 5.
The introduced potential in the GGA+U \cite{ldapu2} formalism is on the Ga $d$
admixture in the ground state charge density contained within the 
sphere of specified mufin tin radius. This does not distinguish
between Ga 3$d$ and 4$d$. Hence since the ground state charge density has
some Ga 4$d$ admixture, the Ga 4$d$ states are also pulled to lower energies.
These interact with the VBM derived As $p$ levels and hence modify the 
bandgap which keeps on increasing with $U$. This is illustrated
in Fig. 6 where we show the valence and conduction band structure 
as $U$ is varied. With $U$ the changes in the valence band are small, 
while the changes in the conduction band are substantial. 
Therefore we clarify for the 
first time that a part of the observed increase in the bandgap with $U$ on the semi core 
states is because of the interaction of the unoccupied Ga 4$d$ states with 
the states comprising the VBM.

The question we asked next was whether we can reduce the component of the 4 $d$ 
states on which the potential is applied. This has been done by varying the 
muffin tin radius about the atom on which the potential is applied and the 
resulting density of states for GaAs are shown in Fig.7.  
For $U$=0 we find that the large RMT result indicates that there is substantial 
Ga $d$ admixture in the valence band. However as the muffin tin radius is 
reduced, this contribution is substantially reduced. Hence for 
large $U$ and large RMT we find the unusual result that the Ga $d$ 
states contribution in the valence band increases. We would 
expect it to decrease by virtue of the fact that we are moving the Ga $3d$ 
states deeper. For small RMT however we do not find an increase in the Ga $d$ component 
in the valence band. The modifications in the bandgap with RMT are given in 
Table II. The changes are small for small RMT, though the bandgap improvement is still 
larger than what is expected from the tight binding model. We examine 
what the modifications of the electronic structure are as a function of $U$ with 
a smaller muffin tin radius of 1.5 a.u.
Aligning the different $U$ calculations with respect to the anion 1$s$ for 
the same $U$, we find that the relative separtion of the anion core levels 
remain unchanged while those of the Ga core levels are strongly modified. 
This is given in Table III. From this we infer that there is no charge 
transfer taking place between Ga and X. However the charge on the Ga gets 
reorganised between the levels resulting in a change in the relative separations 
of the core levels. 
This observation is consistent with the fact that there are modifications of the 
charge on the Ga atoms, while that on the anions is essentially unchanged. 
Hence as a function of $U$, there are changes in the Ga-$d$-anion-$p$ interactions 
which results in substantial modifications of the bandgap. However there is no 
change in the effective charge transfer between the anion and cations sites 
as function of $U$. Constructing a superlattice of the form $(GaX)_{5}/(Ga(U)X)_{5}$ 
we determine the valence band and conduction band offsets between the $U$=0 and 
finite $U$ results. Here the value $U$ has been chosen so that we obtain agreement 
in the position of the experimental Ga 3$d$ positions. These are shown in 
Fig. 8. We find that a $U$ on the semicore state shifts the conduction band 
minimum as well as the valence band maximum to different extends. These conclusions 
are similar to those obtained by Janotti et al. \cite {chris}.

In contrast to the LAPW method where the region over which the GGA+U correction 
is applied is controlled by the choice of the muffin tin radius, such a freedom 
does not exist in pseudopotential calculations. We have calculated the change 
in the bandgap for the GaN, GaP and GaAs within a pseudopotential approach (Table IV). 
The results are similar to our large RMT results.

In Fig. 9 
we have plotted the variation of the approximate position of the
Ga 3$d$ states with respect to the VBM
as a function of $U$. The behaviour is almost
linear and we use this plot to determine the value of $U$ that will bring
about agreement with experiment for the Ga 3$d$ states. Hence a $U$ of
approximately 11, 14 and 13 eV, with a muffin tin radius of 1.5 a.u, for GaN, GaP and GaAs
respectively is found to be sufficient to bring about
agreement with experiment.
Similar range of $U$ was required to get the position of the $d$ states
correct in ZnO \cite{clas-zno}.
We would however like to point out that
GGA+U is
basis set dependent and therefore the $U$ value (for obtaining a certain
effect)
differs between different computational methods,
different choice of muffin-tin radius, etc.

In Fig. 10  we have plotted the variation in the band structure of the 
valence band as well as the conduction band region for GaAs as a function of $U$ for 
a smaller muffin tin radius of 1.5 a.u.
The conduction band minimum has been defined as the zero of the energy scale. There are $k$
dependent changes in the electronic structure that one finds. 
A popular method used in the correction of the band-gap of semiconductors
is one in which all the states are rigidly shifted \cite{scissor}. 
This method results
in a constant shift of a band at all $k$ points, which is indeed 
not the case as evident from Fig. 10. Hence a $k$-dependent self energy is
intrinsic to the GGA+U approach which involves a self consistent solution
of the Kohn-Sham Hamiltonian in the presence of the additional potential.
This has been pointed out by Persson and Mirbt \cite{bjphy}
who explain that LDA underestimates the $\Gamma$-point electron $m_{c}$ and light-hole  
$m_{lh}$ masses due to a too strong k-dependent bonding-antibonding interaction; the  
heavy-hole $m_{hh}$ and spin-orbit split off $m_{so}$ masses are in general less affected
by this interaction.
This LDA failure is very apparant for GaAs where LDA produces a very small bandgap, and  the
corresponding LDA average masses (see Ref. \cite{persson2}  for definition of the geometric average masses) are
$m_{c}$ = 0.01$m_0$, $m_{hh}$ = 0.49$m_0$, $m_{lh}$ = 0.02$m_0$, and $m_{so}$ = 0.07$m_0$,
whereas the experimental values \cite{expt-mass} are
$m_{c}$ = 0.07$m_0$, $m_{hh}$ = 0.53-0.59$m_0$, $m_{lh}$ = 0.08$m_0$, and $m_{so}$ = 0.13$m_0$.
The k-dependent modifications in the GGA+U formalism result in an increase of the bandgap and also a
weakening of the bonding-antibonding interaction, thereby improving the effective masses.
Using the LAPW method with $U$ = 9 eV we obtain
$m_{c}$ = 0.04$m_0$, $m_{hh}$ = 0.62$m_0$, $m_{lh}$ = 0.05$m_0$, and $m_{so}$ = 0.12$m_0$
which agree much better with the measured values.

{\bf IV. Conclusion}

To conclude, we have examined  a popular and computationally inexpensive
method of correcting for the bandgap underestimation in first-principle
calculations. For small values of $U$ one finds that the changes in the valence band 
are confined to the region around the $\Gamma$ point. Suprisingly there are changes 
induced in the unoccupied states as well, as a function of $U$ on the semi-core states. 
This has been pointed out as arising from the fact that the GGA+U potential is applied on 
the $d$ admixture in the ground state charge density. This can be reduced by choosing 
a smaller muffin tin radius for the atom to which GGA+U corrections are applied.
Another component to the bandgap increase is from modified screening of the valence 
states as the Ga $d$ semi-core states are pushed deeper into the 
valence band.

PM thanks DAE-BRNS, India for financial support under 
project number 2005/37/8/BRNS.
We all thank the Swedish Research Council (VR), 
and the Swedish Foundation for International Cooperation in Research and Higher
Education (STINT). 
\renewcommand
\newpage


\begin{figure}
\includegraphics[width=5.5in,angle=270]{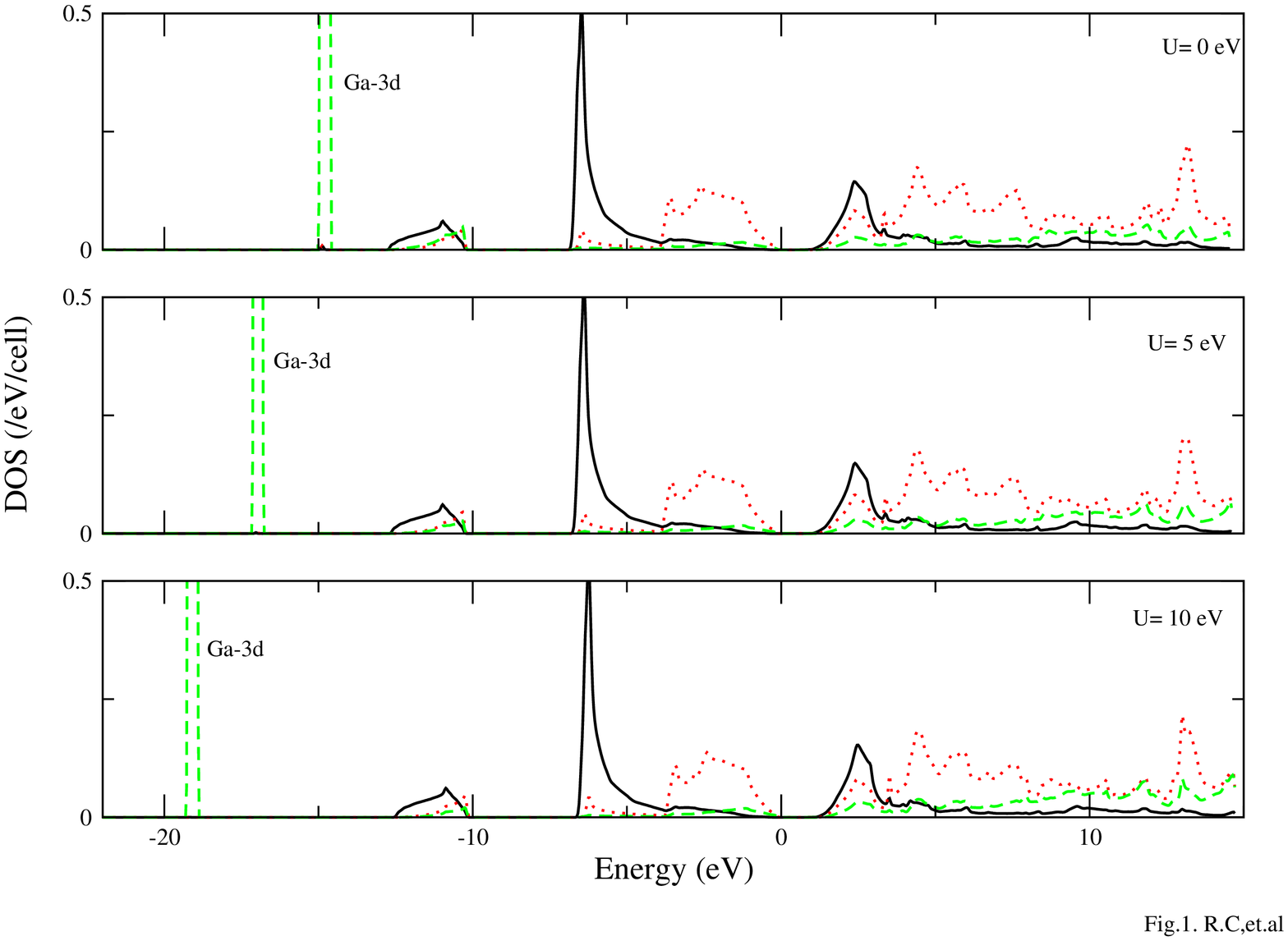}
\caption{ The Ga - $s$ (black solid lines), $p$ (red dotted lines) and $d$ 
(green dashed lines) 
projected density of states for $U$=0, 5 and 10 eV on the Ga 3$d$ states 
for GaAs. The
zero of energy corresponds to the valence band maximum.}
\end{figure}

\begin{figure}
\includegraphics[width=5.5in,angle=270]{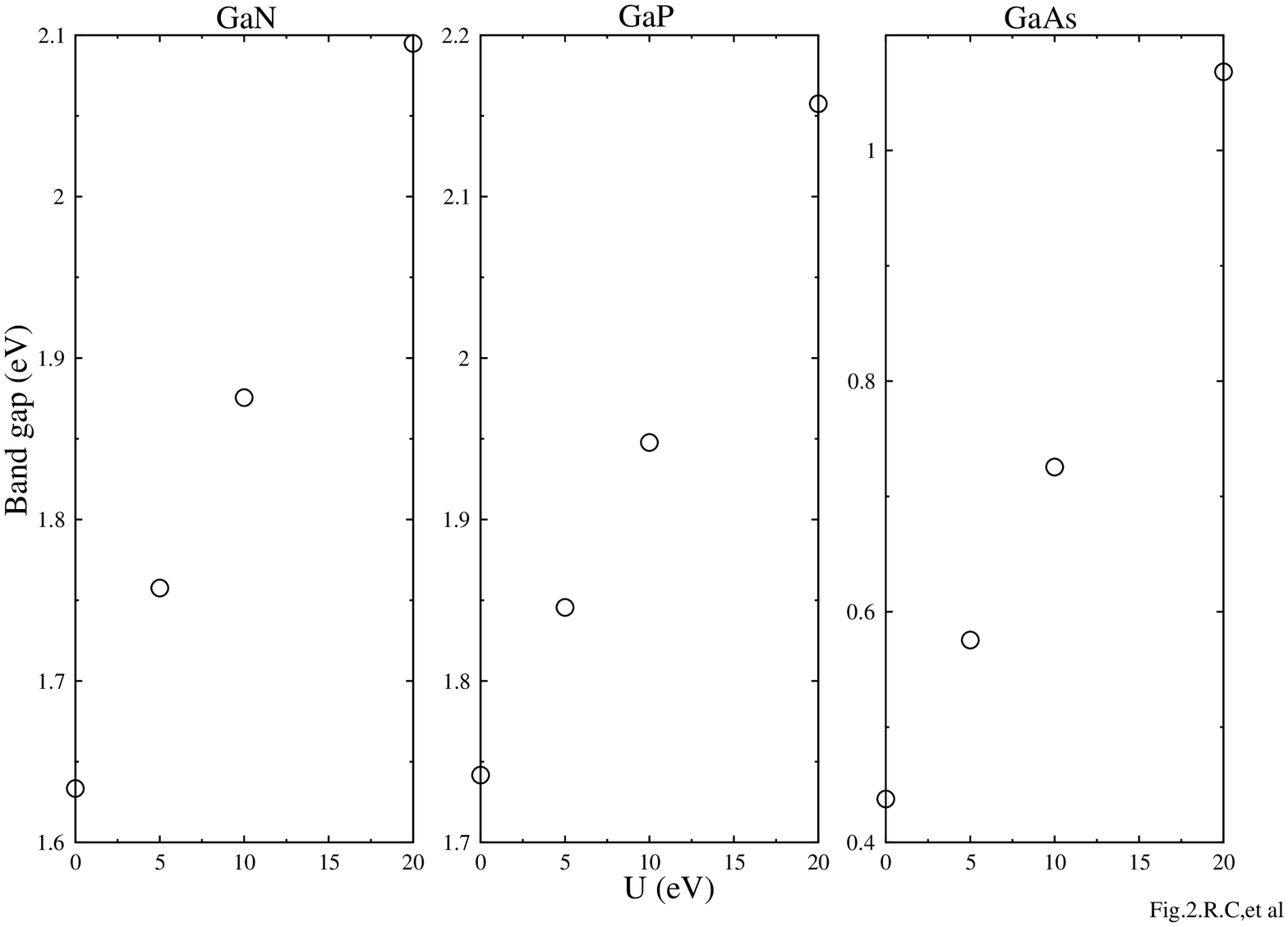}
\caption{ Variation of bandgap as a function of $U$ for GaN, GaP and GaAs.}
\end{figure}

\begin{figure}
\includegraphics[width=5.5in,angle=270]{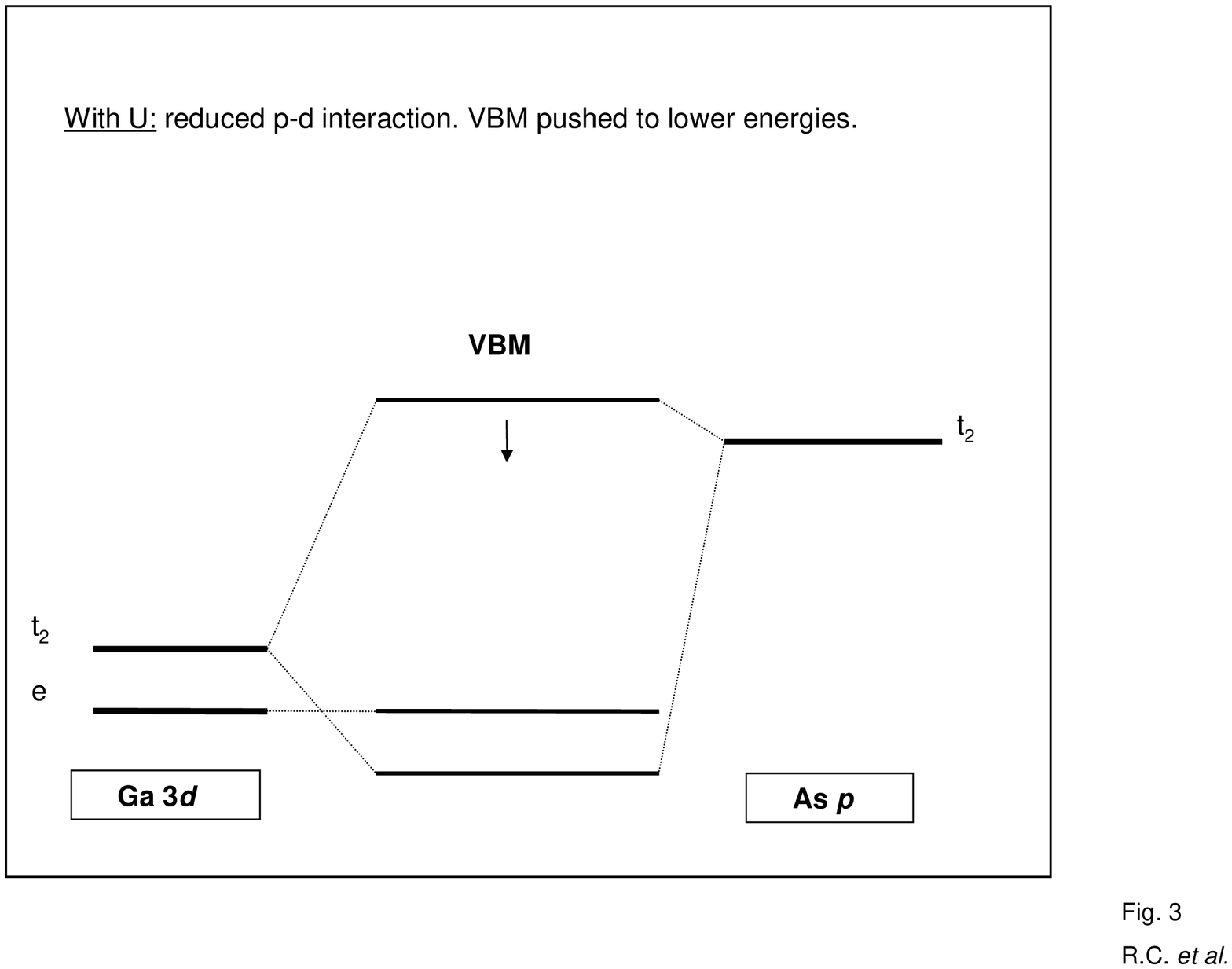}
\caption{ Schematic energy level diagram for the interaction of the
Ga 3$d$ states with the As $p$ states. The arrow
denotes the movement of the VBM with increasing $U$ on the semi core Ga 3$d$
states.}
\end{figure}

\begin{figure}
\includegraphics[width=5.5in,angle=270]{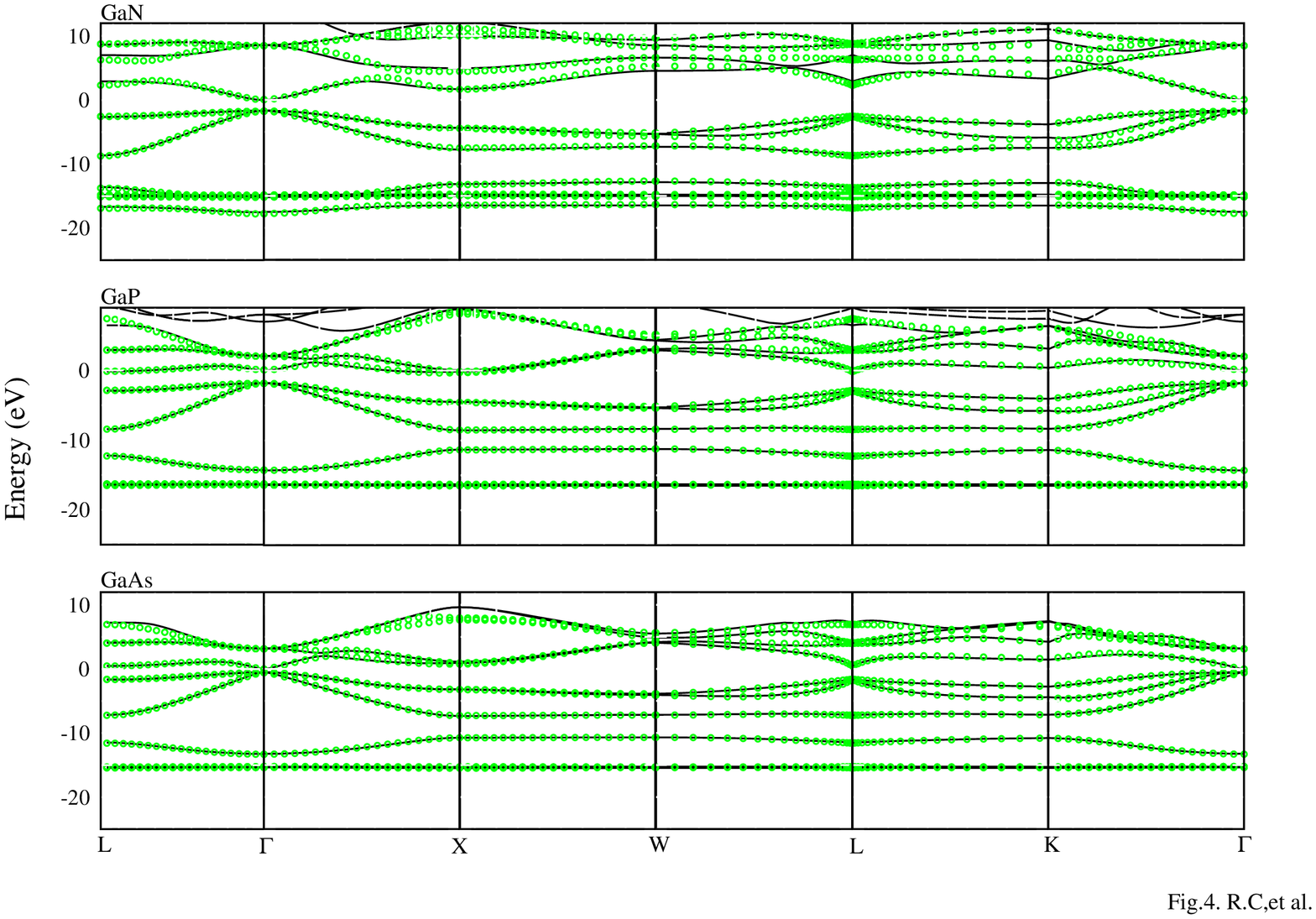}
\caption{ Comparisons of the band dispersions obtained for
GaN, GaP and GaAs, from
ab-initio (solid lines) calculations and from TB fitting (open circles).}

\end{figure}

\begin{figure}
\includegraphics[width=5.5in,angle=270]{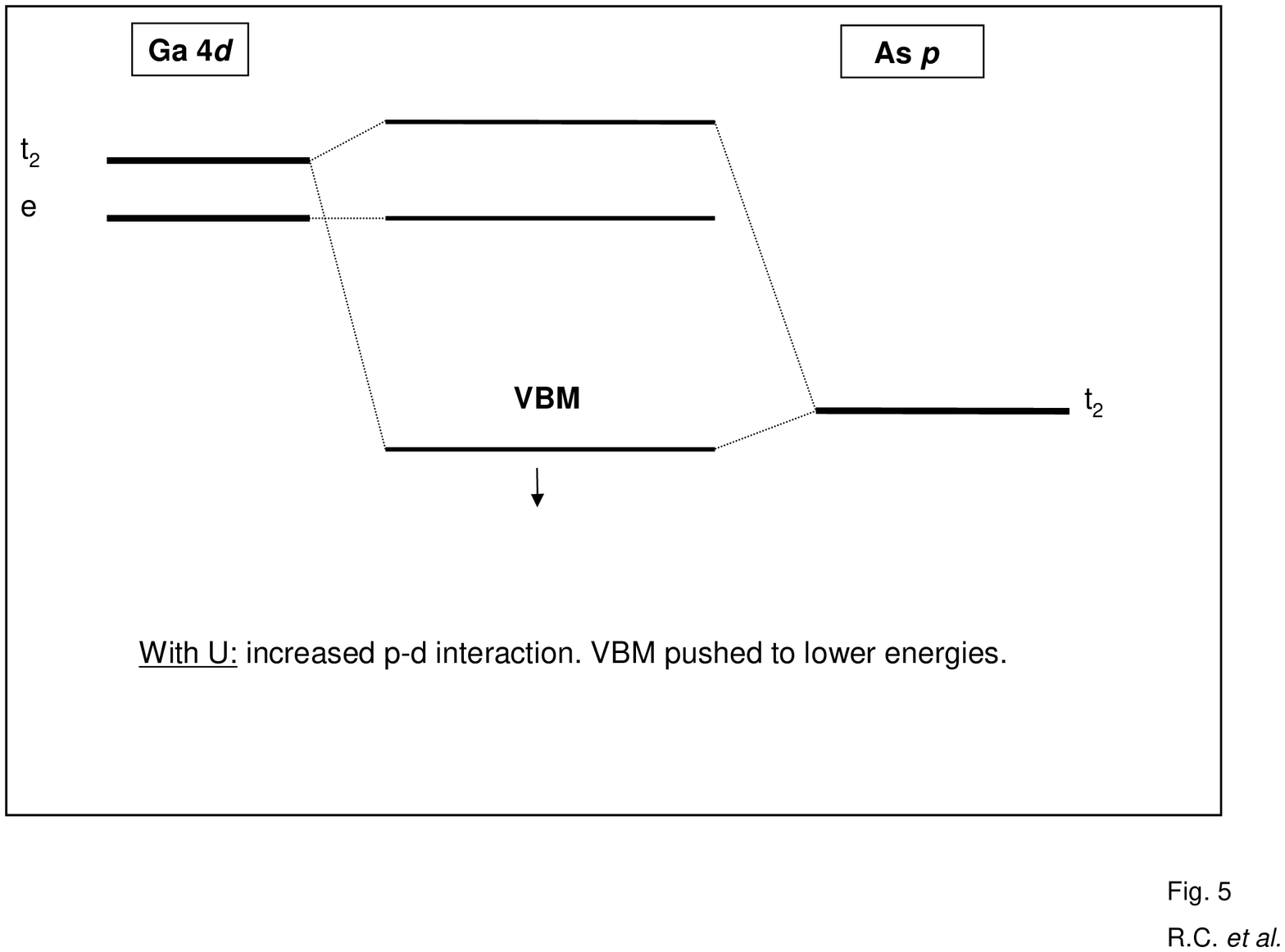}
\caption{ Schematic energy level diagram for the interaction of the
Ga 4$d$ states with the As $p$ states. The arrow
denotes the movement of the VBM with increasing $U$ on the semi core Ga 3$d$
states.}
\end{figure}

\begin{figure}
\includegraphics[width=5.5in,angle=270]{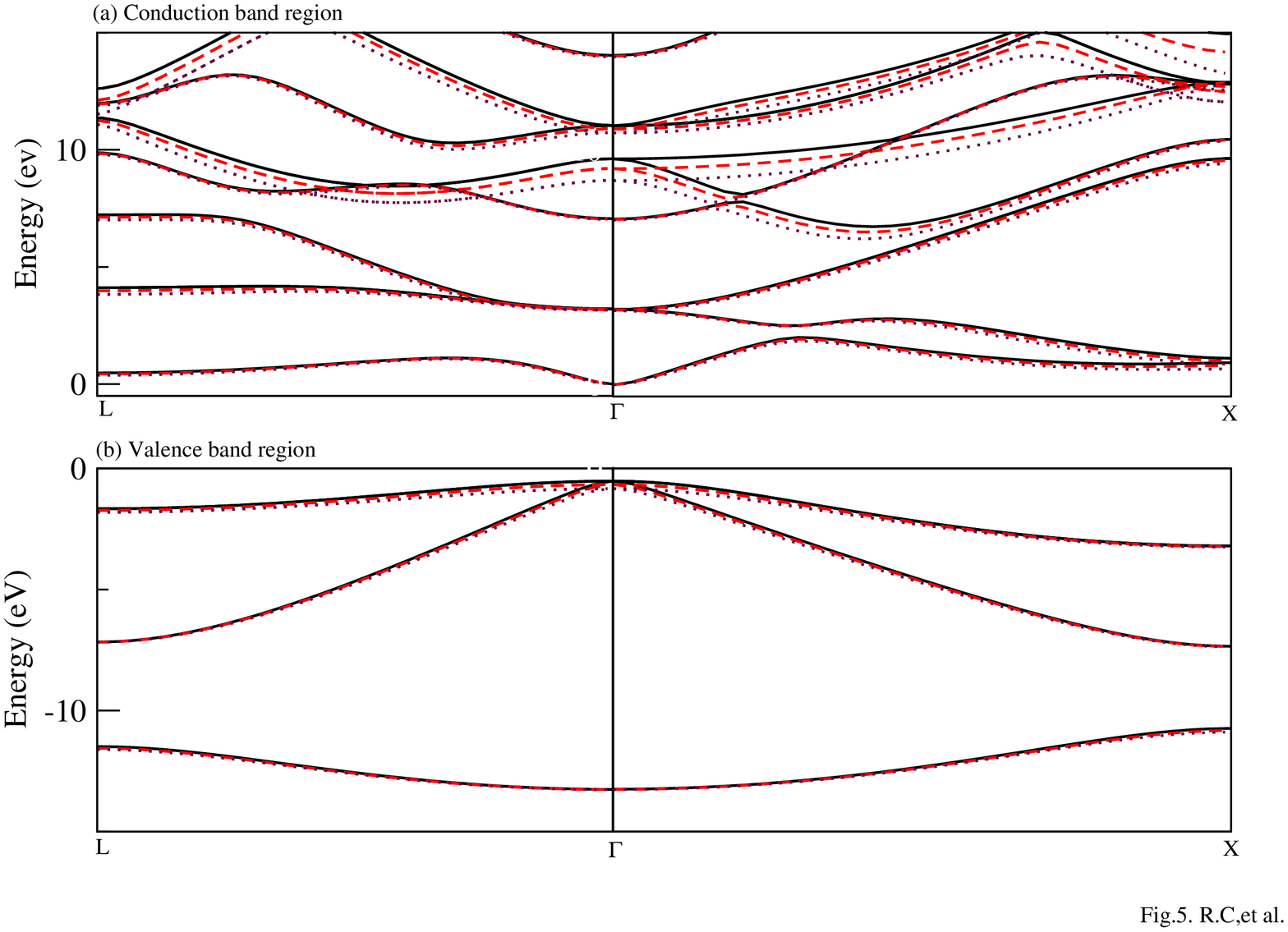}
\caption{ Near $\Gamma$ point view of the
band dispersions obtained for the GaAs
for the $U$ values 0(black solid lines), 5(red dashed lines) and
10(brown dotted lines) eV
(a) in the conduction band region and (b) in the valence band region. The
graphs have been aligned with respect to the As 1s level for each $U$ value.
(RMT of 2.3 a.u for Ga).}
\end{figure}

\begin{figure}
\includegraphics[width=5.5in,angle=270]{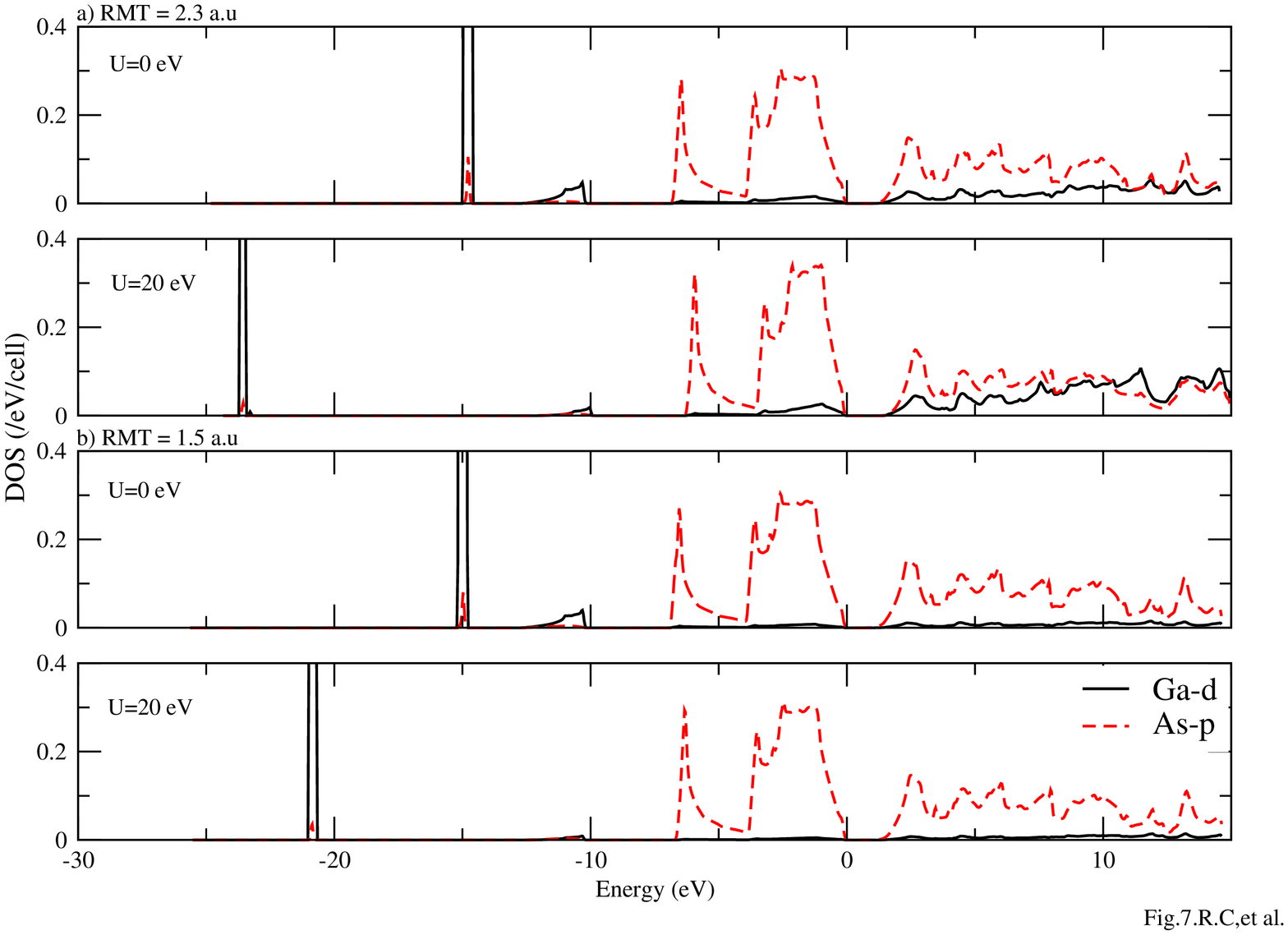}
\caption{The Ga $d$ (black solid lines) and As $p$ (red dashed lines)
projected density of states for $U$=0 and 20 eV on the Ga 3$d$ states
for GaAs with (a) RMT= 2.3 and (b) 1.5 a.u. The zero of energy
corresponds to the valence band maximum. }
\end{figure}

\begin{figure}
\includegraphics[width=5.5in,angle=270]{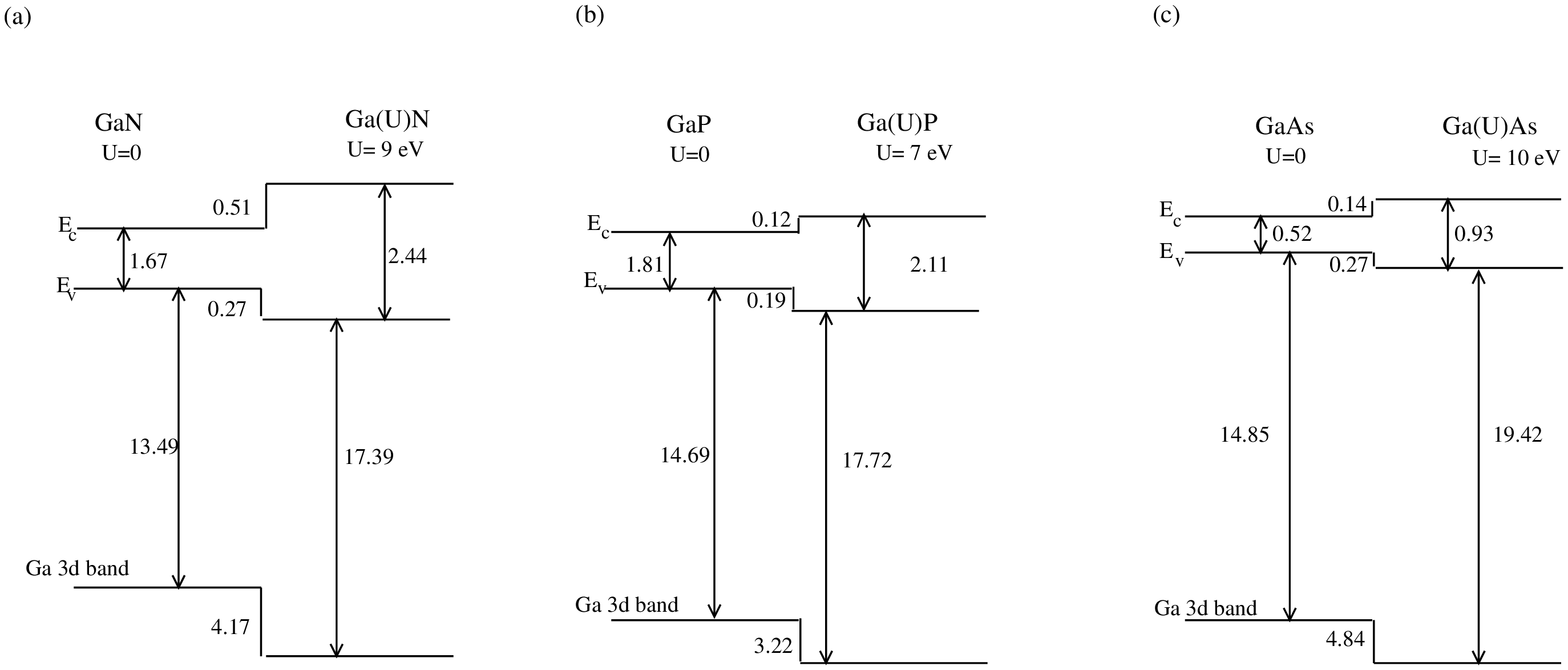}
\caption{ Calculated band offsets at the hypothetical 
interfaces (a) GaN/Ga(U)N (b) GaP/Ga(U)P and (c) GaAs/Ga(U)As 
where a $U$ has been introduced only on the Ga $d$ states.
PAW potentials were used for these calculations.}
\end{figure}

\begin{figure}
\includegraphics[width=5.5in,angle=270]{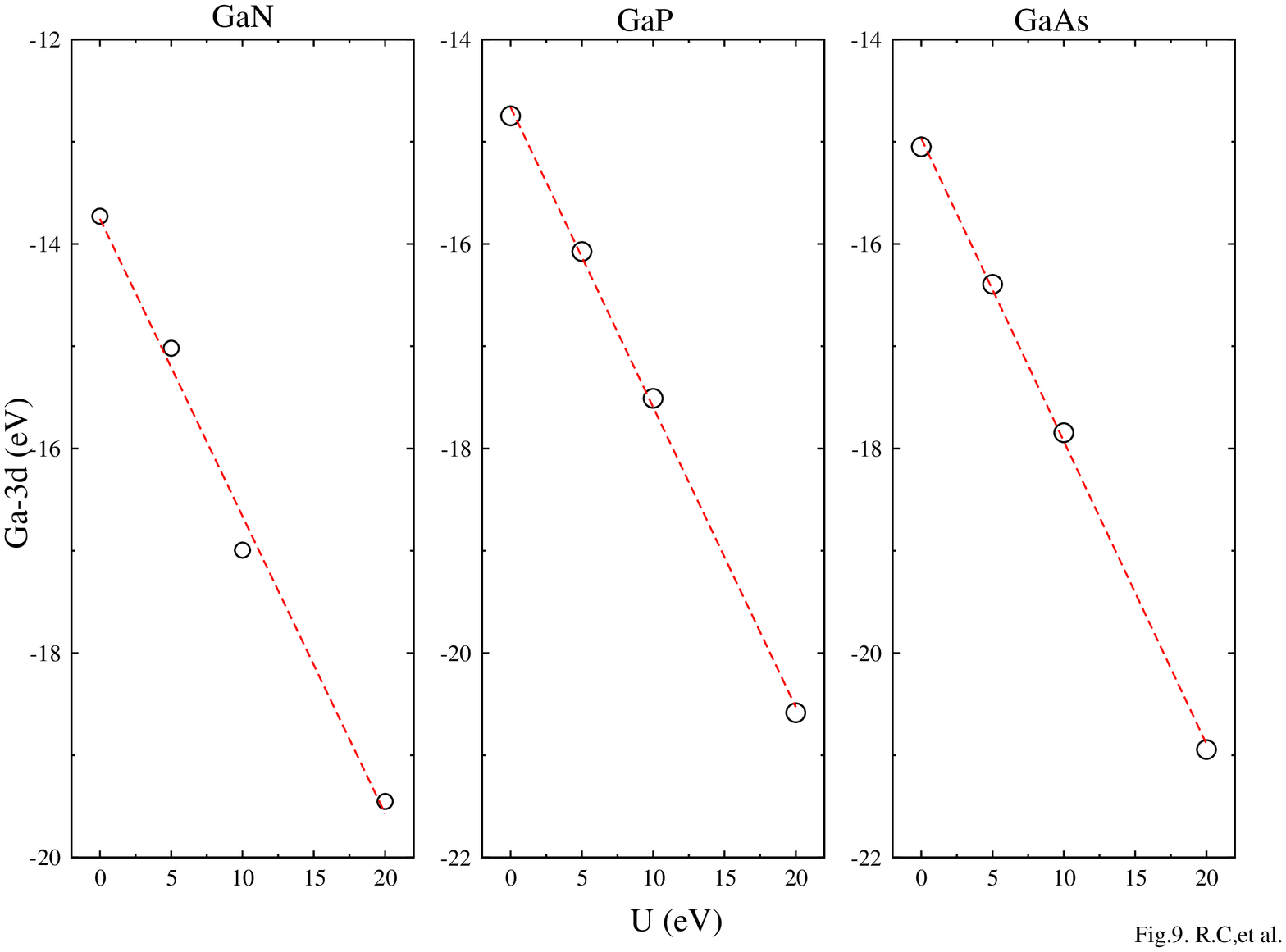}
\caption{ Variation of Ga 3$d$ states position with $U$ for GaN, GaP and GaAs.}
\end{figure}

\begin{figure}
\includegraphics[width=5.5in,angle=270]{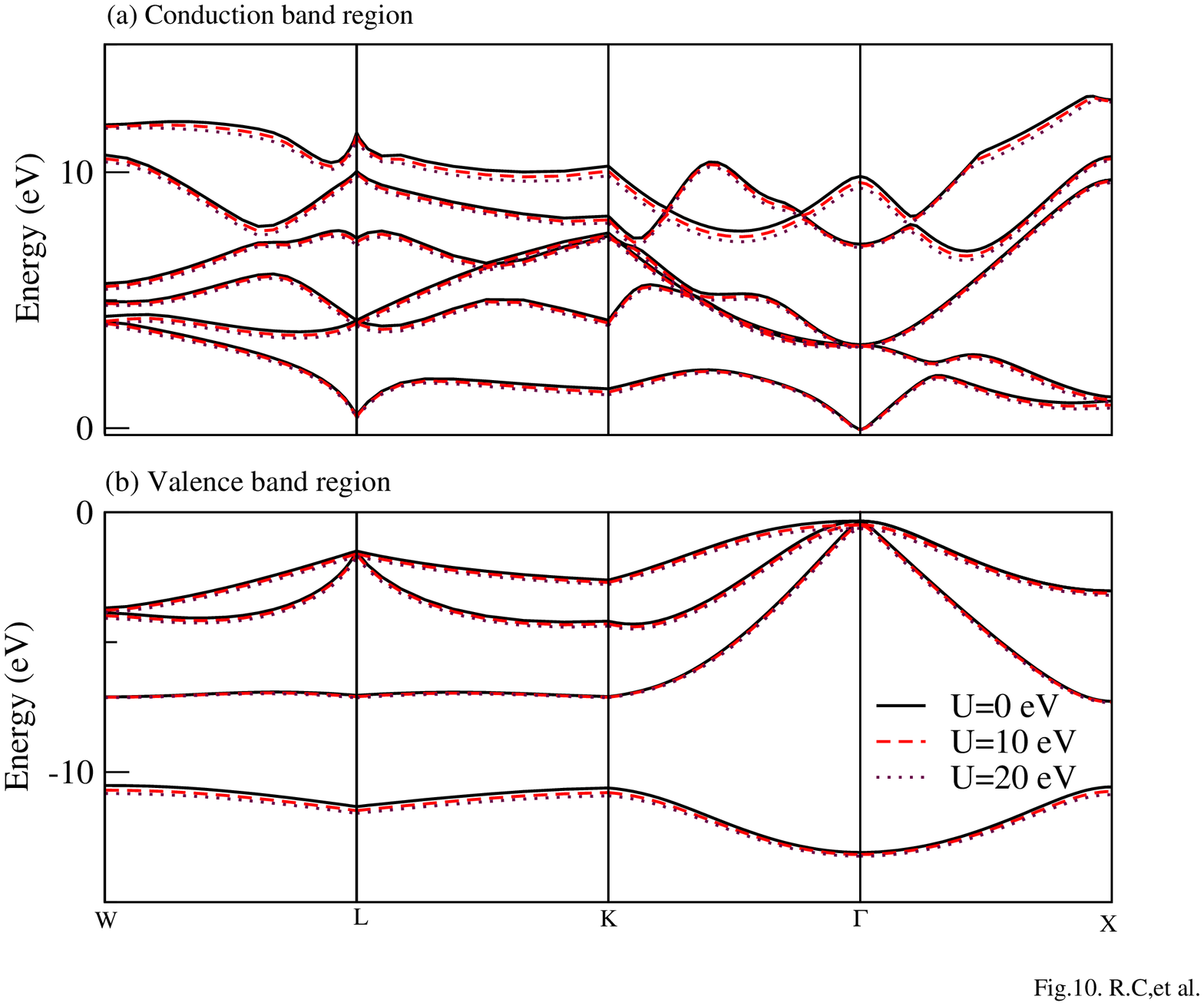}
\caption{ Band dispersions obtained for the GaAs 
for the $U$ values 0(black solid lines), 10(red dashed lines) and 
20(brown dotted lines) eV
(a) in the conduction band region and (b) in the valence band region. The
graphs have been aligned with respect to the As 1s level for each $U$ value.
(RMT of 1.5 a.u for Ga).}
\end{figure}

\newpage

\begin{table}
\caption
{Tight-binding parameters (in eV) obtained from a least-squared-error
fitting procedure for GaX semiconductors. The first six rows contain 
the orbital energies (s$_{c}$, p$_{c}$,..,d$_{a}$), where the main letter denotes orbital type (s,p,d) 
and the subscript denotes the cation (c) or anion (a), followed by the Slater Koster 
parameters. The VBM is set to the zero of the energy.}
\begin{tabular}{l|c|c|c}
\hline\hline
  & GaN & GaP & GaAs \\ \hline\hline
s$_{c}$  & 6.696  & 1.82 & 1.01\\
p$_{c}$ & 9.49 & 7.94 & 6.59\\
d$_{c}$ & -13.16& -14.46 & -14.74 \\
s$_{a}$ & -12.97 & -8.59 &  -9.97\\
p$_{a}$ & -0.92 & -0.46 & -0.367\\
d$_{a}$ & - & 7.71 & 6.88 \\
s$_{c}$s$_{a}$$\sigma$ & -1.42  & -1.59 & -1.33\\
s$_{c}$p$_{a}$$\sigma$ & 3.28 & 2.76 & 2.67 \\
s$_{c}$d$_{a}$$\sigma$ & - & -2.07 & -1.72 \\
p$_{c}$p$_{a}$$\sigma$ & 3.17 & 2.85 & 2.94\\
p$_{c}$p$_{a}$$\pi$ & -0.92 & -1.05 &  -0.81\\
p$_{c}$d$_{a}$$\sigma$ & - & -0.62 & -0.50 \\
p$_{c}$d$_{a}$$\pi$ & - & 1.60 & 1.28\\
d$_{c}$d$_{a}$$\sigma$ & - & 0.0 & 0.0\\
d$_{c}$d$_{a}$$\pi$ & - & 0.0 & 0.0 \\
d$_{c}$d$_{a}$$\delta$ & - & 0.0 & 0.0 \\
p$_{c}$s$_{a}$$\sigma$ & -0.28 & -1.54 & -0.75\\
d$_{c}$s$_{a}$$\sigma$ & -0.77 & 0.0 & 0.0\\
d$_{c}$p$_{a}$$\sigma$ & 1.08 & 0.636 & 0.568\\
d$_{c}$p$_{a}$$\pi$ & -0.01 & -0.224 & -0.318\\
s$_{c}$s$_{c}$$\sigma$ & -0.59 & -0.31 &  -0.26 \\
s$_{c}$p$_{c}$$\sigma$ & 0.69 & 0.46 & 0.05\\
s$_{c}$d$_{c}$$\sigma$ & -0.06 & -0.22 & -0.16 \\
p$_{c}$p$_{c}$$\sigma$ & 1.49 & 0.11 &  0.28\\
p$_{c}$p$_{c}$$\pi$ & 0.0 & -0.04 & -0.23\\
p$_{c}$d$_{c}$$\sigma$ & 0.0 & 0.0 & 0.0\\
p$_{c}$d$_{c}$$\pi$ & 0.0 & 0.03 & 0.0\\
d$_{c}$d$_{c}$$\sigma$ & -0.06 & -0.01 & 0.0\\
\end{tabular}
\end{table}
\newpage
\begin{table}
\begin{tabular}{l|c|c|c}
  & GaN & GaP & GaAs \\ \hline\hline
d$_{c}$p$_{a}$$\pi$ & -0.01 & -0.224 & -0.318\\
s$_{c}$s$_{c}$$\sigma$ & -0.59 & -0.31 &  -0.26 \\
s$_{c}$p$_{c}$$\sigma$ & 0.69 & 0.46 & 0.05\\
s$_{c}$d$_{c}$$\sigma$ & -0.06 & -0.22 & -0.16 \\
p$_{c}$p$_{c}$$\sigma$ & 1.49 & 0.11 &  0.28\\
p$_{c}$p$_{c}$$\pi$ & 0.0 & -0.04 & -0.23\\
p$_{c}$d$_{c}$$\sigma$ & 0.0 & 0.0 & 0.0\\
p$_{c}$d$_{c}$$\pi$ & 0.0 & 0.03 & 0.0\\
d$_{c}$d$_{c}$$\sigma$ & -0.06 & -0.01 & 0.0\\
d$_{c}$d$_{c}$$\pi$ & 0.02& 0.01 & 0.01\\
d$_{c}$d$_{c}$$\delta$ & 0.0 & 0.0 & 0.0 \\
s$_{a}$s$_{a}$$\sigma$ & -0.06 & 0.0 & -0.01\\
s$_{a}$p$_{a}$$\sigma$ & 0.2 & 0.0 & 0.07\\
s$_{a}$d$_{a}$$\sigma$ & - & -0.15 & -0.13\\
p$_{a}$p$_{a}$$\sigma$ & 0.34 & 0.23 & 0.31\\
p$_{a}$p$_{a}$$\pi$ & -0.05 & -0.05 & -0.04\\
p$_{a}$d$_{a}$$\sigma$ & - & -0.24 & -0.31\\
p$_{a}$d$_{a}$$\pi$ & - & 0.15 & 0.13\\
d$_{a}$d$_{a}$$\sigma$ & - & -0.98 & -0.84\\
d$_{a}$d$_{a}$$\pi$ & - & 0.51 & 0.43 \\
d$_{a}$d$_{a}$$\delta$ & - & -0.1 & -0.036\\
\hline
\end{tabular}
\end{table}

\newpage

\begin{table}
\caption
{Bandgap change ($E_g$(U=20)-$E_g$(U=0)) as a function of muffin tin radii.}
\begin{tabular}{l|c|c}
\hline\hline
  & RMT (a.u) & Bandgap change (eV) \\ \hline\hline
GaN  & 1.8 &  0.46 \\
GaN  & 1.5 &  0.34 \\
GaP  & 2.0 &  0.42 \\
GaP  & 1.5 &  0.29 \\
GaAs & 2.3 &  0.63 \\
GaAs & 1.5 &  0.29 \\
\hline
\end{tabular}
\end{table}

\begin{table}
\caption
{The relative shifts on the Ga and X (anion) core levels (Ry.) with respect to the X 1$s$ core level.
(RMT of 1.5 a.u for Ga).}
\begin{tabular}{l|c|c|c|c}
\hline\hline
 &  & Core level & \multicolumn{2}{c} {U (eV)}   \\
\cline{4-5}
 & & & 10 & 20  \\
\hline
\hline
GaN & & &  \\
 &Ga & 1S$_{1/2}$ & -723.683948 & -723.581975 \\
 &Ga & 2S$_{1/2}$ & -64.611804 & -64.501480 \\
 &Ga & 2P$_{1/2}$  & -53.981829 & -53.871907 \\
 &Ga & 2P$_{3/2}$ &-51.947788 & -51.837825 \\
 &Ga & 3S$_{1/2}$ & 17.218648 & 17.295227  \\
 &Ga & 3P$_{1/2}$  & 20.651111 & 20.725591\\
 &Ga & 3P$_{3/2}$ & 20.912098 & 20.985360 \\
 &N & 1S$_{1/2}$ & 0.0 & 0.0 \\
\hline\hline
GaP &  & &  \\
&Ga & 1S$_{1/2}$ & -598.619678 & -598.517630 \\
&Ga & 2S$_{1/2}$ & 60.447143 & 60.558034 \\
&Ga & 2P$_{1/2}$  & 71.077537 & 71.187967 \\
&Ga & 2P$_{3/2}$ & 73.111528 & 73.222006 \\
&Ga & 3S$_{1/2}$ & 142.286256 & 142.362896 \\
&Ga & 3P$_{1/2}$ & 145.719349& 145.793838 \\
&Ga & 3P$_{3/2}$ & 145.980772 & 146.054018 \\
&P & 1S$_{1/2}$ & 0.0  & 0.0 \\
&P & 2S$_{1/2}$ & 140.426058 & 140.424829  \\
&P & 2P$_{1/2}$  & 144.012722 & 144.011600 \\
&P & 2P$_{3/2}$ & 144.080163 & 144.079036 \\
\hline\hline
\end{tabular}
\end{table}
\newpage
\begin{table}
\begin{tabular}{l|c|c|c|c}
\hline\hline
 &  & Core level & \multicolumn{2}{c} {U (eV)}   \\
\cline{4-5}
 & & & 10 & 20  \\
\hline
\hline
GaAs & & &  \\
&Ga & 1S$_{1/2}$ & 109.385718 & 109.544887  \\
&Ga & 2S$_{1/2}$ & 768.448458 & 768.620852 \\
&Ga & 2P$_{1/2}$ & 779.078965 & 779.250701 \\
&Ga & 2P$_{3/2}$ & 781.112945 & 781.284749 \\
&Ga & 3S$_{1/2}$ & 850.307135 & 850.427101 \\
&Ga & 3P$_{1/2}$ & 853.741405 & 853.858096 \\
&Ga & 3P$_{3/2}$ & 854.003530 & 854.118328 \\
&As & 1S$_{1/2}$ & 0.0  &  0.0\\
&As & 2S$_{1/2}$ & 752.001092 & 752.000418  \\
&As & 2P$_{1/2}$ & 763.627530 & 763.626944 \\
&As & 2P$_{3/2}$ & 766.316562 & 766.315964 \\
&As & 3S$_{1/2}$ & 847.269176 & 847.268283 \\
&As & 3P$_{1/2}$ & 851.175282 & 851.174383 \\
&As & 3P$_{3/2}$ & 851.544909 & 851.544006 \\
\hline\hline
\end{tabular}
\end{table}

\begin{table}
\caption
{Bandgap as a function of $U$ from VASP}
\begin{tabular}{l|c|c}
\hline\hline
  & $U$ (eV) & Bandgap (eV) \\ \hline\hline
GaN  & 0 &  1.633  \\
GaN  & 5 &  1.758  \\
GaN  & 10 & 1.875   \\
GaN  & 20 & 2.095   \\
GaP  & 0 &  1.742 \\
GaP  & 5 &  1.846 \\
GaP  & 10 & 1.948   \\
GaP  & 20 & 2.157   \\
GaAs & 0 &  0.438 \\
GaAs & 5 &  0.576 \\
GaAs & 10 & 0.725   \\
GaAs & 20 & 1.068   \\
\hline
\end{tabular}
\end{table}

\end{document}